\pdfoutput=1

\documentclass[11pt,twoside,a4paper,cmspaper,final,collab]{cms-tdr}
\def\svnVersion{472964P}\def\cmsCernNoTag{CERN-EP-2017-285}\def\cmsCernDate{\today}\def\cmsMessage{Published in Physics Letters B as \href{http://dx.doi.org/10.1016/j.physletb.2018.07.061}{\doi{10.1016/j.physletb.2018.07.061.}}}

\begin{document}\cmsNoteHeader{HIN-16-002}

\hyphenation{had-ron-i-za-tion}
\hyphenation{cal-or-i-me-ter}
\hyphenation{de-vices}

\RCS$Revision: 472131 $
\RCS$HeadURL: svn+ssh://svn.cern.ch/reps/tdr2/papers/HIN-16-002/trunk/HIN-16-002.tex $
\RCS$Id: HIN-16-002.tex 472131 2018-08-17 15:01:10Z rbi $
\newlength\cmsFigWidth
\ifthenelse{\boolean{cms@external}}{\setlength\cmsFigWidth{0.98\columnwidth}}{\setlength\cmsFigWidth{0.65\textwidth}}
\ifthenelse{\boolean{cms@external}}{\providecommand{\cmsLeft}{upper\xspace}}{\providecommand{\cmsLeft}{left\xspace}}
\ifthenelse{\boolean{cms@external}}{\providecommand{\cmsRight}{lower\xspace}}{\providecommand{\cmsRight}{right\xspace}}
\newlength\cmsTabSkip\setlength\cmsTabSkip{2ex}
\newcommand {\pp}    {\ensuremath{\Pp\Pp}\xspace}
\newcommand {\PbPb}  {\ensuremath{\mathrm{PbPb}}\xspace}
\newcommand {\ptg}    {\ensuremath{\pt^\gamma}\xspace}
\newcommand {\ptj}    {\ensuremath{\pt^\text{jet}}\xspace}
\newcommand {\etag}    {\ensuremath{\eta^\gamma}\xspace}
\newcommand {\etaj}    {\ensuremath{\eta^\text{jet}}\xspace}
\newcommand {\npart}  {\ensuremath{\langle N_{\text{part}} \rangle}\xspace}
\providecommand{\HYDJET} {{\textsc{hydjet}}\xspace}
\providecommand{\pythiahydjet} {\PYTHIA{}+\HYDJET{}\xspace}
\newcommand{\jewel} {\textsc{jewel}\xspace}
\newcommand {\ptGen}    {\ensuremath{\pt^\text{gen}}\xspace}
\newcommand {\ptReco}    {\ensuremath{\pt^\text{reco}}\xspace}
\newcommand {\ptjet}    {\ensuremath{\pt^\text{jet}}\xspace}
\providecommand{\rjg}{\ensuremath{R_{\mathrm{j}\gamma}}\xspace}
\providecommand{\xjg}{\ensuremath{x_{\mathrm{j}\gamma}}\xspace}
\providecommand{\ptg}{\ensuremath{p_{\mathrm{T},\gamma}}\xspace}
\providecommand{\dphijg}{\ensuremath{\Delta\phi_{\mathrm{j}\gamma}}\xspace}
\providecommand{\avexjg}{\ensuremath{\langle x_{\mathrm{j}\gamma} \rangle}\xspace}
\newcommand{\iaa}{\ensuremath{\mathrm{I_\mathrm{AA}^\text{jet}}}\xspace}
\newcommand {\phig}    {\ensuremath{\phi^\cPgg}\xspace}
\newcommand {\phij}    {\ensuremath{\phi^\text{jet}}\xspace}
\newcommand {\sumIso}{\ensuremath{\text{SumIso}^{\mathrm{UE - sub}}}\xspace}
\newcommand {\sumIsoGen}{\ensuremath{\text{SumIso}}\xspace}

\cmsNoteHeader{HIN-16-002}

\title{Study of jet quenching with isolated-photon+jet correlations in \PbPb and \pp collisions at $\sqrt{s_{_{\mathrm{NN}}}} = 5.02\TeV$}

\date{\today}

\abstract{
Measurements of azimuthal angle and transverse momentum (\pt) correlations of
isolated photons and associated jets are reported for \pp and \PbPb collisions
at $\sqrtsNN = 5.02\TeV$. The data were recorded with the CMS detector at the
CERN LHC. For events containing a leading isolated photon with $\ptg > 40\GeVc$
and an associated jet with $\ptjet > 30\GeVc$, the photon+jet azimuthal
correlation and \pt imbalance in \PbPb collisions are studied as functions of
collision centrality and $\ptg$. The results are compared to \pp reference data
collected at the same collision energy and to predictions from several
theoretical models for parton energy loss. No evidence of broadening of the
photon+jet azimuthal correlations is observed, while the ratio $\ptjet/\ptg$
decreases significantly for \PbPb data relative to the \pp reference. All
models considered agree within uncertainties with the data. The number of
associated jets per photon with $\ptg > 80\GeVc$ is observed to be shifted
towards lower $\ptjet$ values in central \PbPb collisions compared to \pp
collisions.}

\hypersetup{%
pdfauthor={CMS Collaboration},%
pdftitle={Study of jet quenching with isolated photon and jet correlations in PbPb and pp collisions at sqrt(sNN) = 5.02 TeV},%
pdfsubject={CMS},%
pdfkeywords={CMS, heavy ion, physics, photon, jet, jet quenching}}

\maketitle

\section{Introduction}

Quantum chromodynamics predicts that in relativistic heavy ion collisions
a state of deconfined quarks and gluons known as the quark-gluon plasma (QGP)
can be formed~\cite{PhysRevLett.34.1353,Karsch:1995sy}. Parton scatterings with
large momentum transfer, which occur very early (${\approx}0.1\unit{fm}/c$)
compared to the timescale of QGP formation (${\approx}1\unit{fm}/c$), provide
tomographic probes of the plasma~\cite{PhysRevD.27.140}. The outgoing partons
interact strongly with the QGP and lose
energy~\cite{Appel:1985dq,Blaizot:1986ma,Gyulassy:1990ye,Wang:1991xy,
Baier:1996sk,Zakharov:1997uu}. This phenomenon, known as ``jet quenching'', has
been observed through measurements of hadrons with high transverse momentum
(\pt)~\cite{Adams:2003kv,Adare:2008qa,Abelev:2012hxa,Aad:2015wga,CMS:2012aa,Khachatryan:2016odn}
and of
jets~\cite{Chatrchyan:2011sx,Aad:2010bu,ATLAS:2014cpa,Adam:2015doa,Khachatryan:2016jfl,Adam:2015ewa,Adamczyk:2016fqm},
both created by the fragmentation of the high-momentum partons.

Since electroweak bosons do not interact strongly with the
QGP~\cite{Aad:2015lcb,Chatrchyan:2012vq,Chatrchyan:2012nt,Chatrchyan:2014csa},
measurements of jets produced in the same hard scattering in conjunction with
these bosons have, in contrast to dijet measurements, a controlled
configuration of the initial hard
scattering~\cite{Kartvelishvili:1995fr,Wang:1996yh,Wang:1996pe}. The
electroweak boson \pt reflects, on average, the initial energy of the
associated parton that fragments into the jet, before any medium-induced energy
loss has occurred~\cite{Dai:2012am,Kang:2017xnc}. At LHC energies, the
production of jets with $\pt > 30\GeVc$ that are associated with electroweak
bosons is dominated by quark fragmentation~\cite{Neufeld:2010fj}. Hence, the
study of correlations in boson-jet events, such as the azimuthal angle ($\phi$)
difference and \pt ratio between the boson and the associated jets, opens the
possibility for in-depth studies of the parton energy loss mechanisms utilizing
theoretically well-controlled initial production processes. These studies also
facilitate the extraction of QGP properties via comparisons with theoretical
models~\cite{Neufeld:2012df,Wang:2013cia,hybrid:2014,hybrid:2015,jewel:2016,Kang:2017xnc}.
Measurements of this kind were first performed in \PbPb collisions at
a nucleon-nucleon center-of-mass energy $\sqrtsNN = 2.76\TeV$ with
isolated-photon+jet events~\cite{Chatrchyan:2012gt} and at 5.02\TeV with \Z-jet
events~\cite{Sirunyan:2017jic} by the CMS Collaboration at the CERN LHC. The
precision of these previous measurements was limited by the available number of
boson-jet pairs.

In the results reported in this paper, the electroweak boson is an isolated
photon, which is selected experimentally by using an isolation requirement,
namely that the additional energy in a cone of fixed radius around the
direction of the reconstructed photon is less than a specified
value~\cite{Aad:2015lcb,Chatrchyan:2012vq}. This restriction suppresses the
background contributions from photons originating from decays of neutral mesons
(``decay photons"), and gives a sample containing mostly prompt photons. Prompt
photons are photons produced directly in the hard scattering process, or
emitted in the fragmentation of a high-\pt parton (``fragmentation photons").
This Letter reports the measurement of correlations of isolated photons and
associated jets in \PbPb and \pp collisions at \sqrtsNN = 5.02\TeV. The \PbPb
and \pp data samples were collected by the CMS experiment in 2015 and
correspond to integrated luminosities of 404\mubinv and 27.4\pbinv,
respectively. The measurement characterizes parton energy loss through the
$\phi$ and \pt correlations between isolated photons and the associated jets.
The azimuthal angle difference $\dphijg = \abs{\phij - \phig}$, the \pt ratio
$\xjg = \ptj/\ptg$ and its average \avexjg, the average number of associated
jets per photon, \rjg, and the ratio of the yield of associated jets in \PbPb
data to \pp data, \iaa, are presented. The results from \PbPb collisions are
compared to those from \pp collisions, with the \pp data serving as a reference
to extract information about the modifications due to the presence of the QGP.

\section{The CMS detector} \label{sec:CMS}

The central feature of the CMS apparatus is a superconducting solenoid of
6\unit{m} internal diameter, providing a magnetic field of 3.8\unit{T}. Within
the solenoid volume are a silicon pixel and strip tracker which measures
charged particles within the pseudorapidity range $\abs{\eta} < 2.5$, a lead
tungstate crystal electromagnetic calorimeter (ECAL), and a brass and
scintillator hadron calorimeter (HCAL), each composed of a barrel and two
endcap sections. The barrel and endcap calorimeters provide $\abs{\eta}$
coverage out to 3. Photon candidates used in this analysis are reconstructed
using the energy deposited in the barrel region of the ECAL, which covers
a range of $\abs{\eta} < 1.48$. Hadron forward (HF) calorimeters extend the
$\abs{\eta}$ coverage of the HCAL to $\abs{\eta} = 5.2$. In \PbPb collisions,
the HF calorimeters are used to determine the centrality of the collisions,
which is related to the impact parameter of the two colliding Pb
nuclei~\cite{Chatrchyan:2011sx}, and the azimuthal angle of maximum particle
density (the event plane)~\cite{Chatrchyan:2012xq}. Muons are detected in
gas-ionization chambers embedded in the steel flux-return yoke outside the
solenoid. A more detailed description of the CMS detector, together with
a definition of the coordinate system used and the relevant kinematic
variables, can be found in Ref.~\cite{Chatrchyan:2008zzk}.

\section{Analysis procedure} \label{sec:analysis}

\subsection{Event selection}
\label{sec:event_cent}

Events containing high-\pt photon candidates are selected by the CMS trigger
system, which consists of a level-1 (L1) and a high-level trigger
(HLT)~\cite{Khachatryan:2016bia}. Events are first selected by requiring an
ECAL transverse energy deposit larger than 21~(20)\GeV during the \PbPb (\pp)
data-taking period. Photon candidates are then reconstructed at the HLT using
the ``island'' clustering algorithm~\cite{EGM-14-001,Chatrchyan:2012vq}, which
is applied to energy deposits in the ECAL. The HLT selection efficiency was
determined in data and was found to be greater than 98\% for events containing
a photon with $\ptg > 40\GeVc$ and $\abs{\etag} < 1.44$ reconstructed offline.
The $\etag$ interval of the photons used in this analysis is restricted to the
barrel region of the ECAL, which has the best performance in terms of photon
reconstruction and triggering and has the lowest rate of misreconstructed
tracks.

A pure sample of inelastic hadronic \pp and \PbPb collisions is obtained with
further offline selection criteria applied to the triggered
events~\cite{Khachatryan:2010us,Chatrchyan:2011sx}. Notable among these,
a reconstructed event vertex and at least three (one) calorimeter towers in the
HF on each side of the interaction point with energy $>$3\GeV are (is) required
in the \PbPb (\pp) analysis. Events with spurious energy depositions in the
HCAL (\ie, sporadic uncharacteristic noise and signals from malfunctioning
calorimeter channels) are rejected by established algorithms that flag such
events, to remove possible contamination of the jet
sample~\cite{Chatrchyan:2009hy}. Events with multiple collisions have
a negligible effect on the measurement since the average number of collisions
per bunch crossing is around 0.9 for \pp collisions, and less than 0.01 for
\PbPb collisions.

In \PbPb collisions, the centrality measurement is based on percentiles of the
distribution of the total energy measured in both HF calorimeters. The event
centrality observable corresponds to the fraction of the total inelastic
hadronic cross section, starting at 0\% for the most central collisions, \ie,
those with the smallest impact parameter and the largest nuclear
overlap~\cite{Chatrchyan:2011sx}.

\subsection{Jet reconstruction} \label{sec:jetrecon}

Offline jet reconstruction is performed using the CMS particle-flow (PF)
algorithm~\cite{Sirunyan:2017ulk}. By combining information from all
subdetector systems, the PF algorithm identifies final-state particles in an
event, classifying them as electrons, muons, photons, charged hadrons, or
neutral hadrons. To form jets, these PF objects are clustered using the
anti-\kt sequential recombination algorithm provided in the \FASTJET
framework~\cite{Cacciari:2008gp,Cacciari:2011ma}. A small jet radius parameter
of $R = 0.3$ is chosen to minimize the effects of heavy ion background
fluctuations ( $\sim10$\GeV in central \PbPb collisions) and for consistency
with the previous measurement at 2.76\TeV~\cite{Chatrchyan:2012gt}.

For the \PbPb data, the underlying background from soft collisions (\ie, the
underlying event, UE) is subtracted during jet reconstruction by employing the
iterative algorithm described in Ref.~\cite{Kodolova:2007hd}, using the same
implementation as in the \PbPb analysis of Ref.~\cite{Chatrchyan:2011sx}. In
\pp collisions, jets are reconstructed without UE subtraction. For \pp and
\PbPb samples, the reconstructed jet energies are corrected to the energies of
final-state particle jets using a factorized multistep
approach~\cite{Chatrchyan:2011ds}. The corrections are derived using simulated
dijet and photon+jet events generated with the
\PYTHIA8.212~\cite{Sjostrand:2007gs} (CUETP8M1 tune~\cite{Khachatryan:2015pea})
Monte Carlo (MC) event generator which, for the case of \PbPb corrections, are
embedded into a simulated underlying background event from
\HYDJET1.9~\cite{Lokhtin:2005px}. The background simulation is tuned to
reproduce the observed charged-particle multiplicity and \pt spectrum in \PbPb
data. Reconstructed jets are required to have $\abs{\etaj}<1.6$ and corrected
$\ptj > 30\GeVc$, to ensure that the jet reconstruction efficiency and energy
resolution (JER) are well understood, \ie, results from data are in agreement
with expectations from MC.

\subsection{Photon reconstruction} \label{sec:photon_reconstruction}

Photon candidates are reconstructed from clusters of energy deposited in the
ECAL. The ``hybrid'' algorithm used for the analysis in \pp collisions is
detailed in Ref.~\cite{EGM-14-001}, while the description of the island
clustering algorithm optimized for high-multiplicity \PbPb collisions can be
found in Ref.~\cite{Chatrchyan:2012vq}.

In order to reduce electron contamination, photon candidates are discarded if
the differences in pseudorapidity and azimuthal angle between the photon
candidate and any electron candidate track with $\pt > 10$\GeVc are less than
0.02 and 0.15 radians, respectively~\cite{Chatrchyan:2012vq}. These matching
windows are conservative choices based on the detector angular resolution. The
relatively large azimuthal angle window allows for the curvature of the
electron trajectories. Anomalous signals caused by the interaction of highly
ionizing particles directly with the silicon avalanche photodiodes used for the
ECAL barrel readout are removed using the prescription given in
Ref.~\cite{Chatrchyan:2012vq}. The energy of the reconstructed photons is
corrected to account for the effects of the material in front of the ECAL and
for the incomplete containment of the shower energy. For \PbPb data, an
additional correction is applied to account for energy contamination from the
UE. The magnitude of the combined energy correction for isolated photons varies
from 0 to 10\%, depending on the centrality of the collision and \ptg. The
corrections are obtained from simulated \PYTHIA and \pythiahydjet photon
events.

Similar to Ref.~\cite{Khachatryan:2010fm}, a generator-level photon candidate
is considered isolated if the \pt sum of final-state generated particles,
excluding neutrinos, in a cone of radius $\Delta R = \sqrt{\smash[b]{(\Delta
\eta)^2 + (\Delta \phi)^2}}=0.4$ around the direction of the candidate,
\sumIsoGen, is less than 5\GeVc. For a reconstructed photon candidate, the
corresponding isolation variable, \sumIso, is calculated with respect to the
centroid of the cluster, not including the \pt of the cluster and after
correcting for the UE (only in \PbPb collisions), and is required to be less
than 1\GeVc. The isolation criterion for reconstructed photons is tighter than
for generated photons to minimize the impact of UE fluctuations in \PbPb
collisions, where a downward fluctuation in the UE could inadvertently allow
a nonisolated photon candidate to pass the isolation criteria. A systematic
uncertainty is assigned to account for the effect of this difference on the
final observables, as detailed in Section~\ref{sec:systematics}.

Imposing the isolation requirement suppresses the background contributions from
fragmentation and decay photons, resulting in a sample enriched in isolated
prompt photons. The dominant remaining backgrounds for isolated photon
candidates are ECAL showers initiated by isolated hadrons, and real photons
that are decay products of isolated neutral mesons, \eg, $\pi^{0}$, $\eta$, and
$\omega$. The hadron-induced showers are rejected using the ratio of HCAL over
ECAL energy inside a cone of radius $\Delta R = 0.15$ around the photon
candidate, $H/E$. Only photon candidates with $H/E < 0.1$ are selected for this
analysis. The decay photons can be significantly reduced using a cut on the
shower shape, a measure of how energy deposited in the ECAL is distributed in
$\phi$ and $\eta$~\cite{Khachatryan:2010fm}, as discussed in
Section~\ref{sec:photon_jet_selection}. The efficiencies of these criteria in
selecting photons are extracted from simulations as a function of \ptg and
corrected for in collision data.

\subsection{Photon+jet pair selection}
\label{sec:photon_jet_selection}

To form photon+jet pairs, the highest \pt isolated photon candidate that passes
the selection criteria is paired with all jets in the same event. The
combinatorial background in \PbPb collisions, which includes misidentified jets
that arise from UE fluctuations, as well as jets from multiple hard
parton-parton scatterings in the same collision, needs to be subtracted in
order to study the energy loss effects on the jets produced in the same hard
scattering as the photon. This background subtraction is performed by
correlating each leading isolated photon candidate with reconstructed jets
found in 40 different events, randomly selected from minimum bias \PbPb data
such that the event centrality, the interaction vertex position along the beam
axis, and the event plane, are within 5\%, 5\cm, and $\pi/10$, respectively, of
those from the signal event. The values were optimized such that the
statistical uncertainty due to the subtraction is negligible compared to the
statistical uncertainty of the photon sample.

\begin{figure*}[ht]
  \centering
    \includegraphics[width=0.98\textwidth]{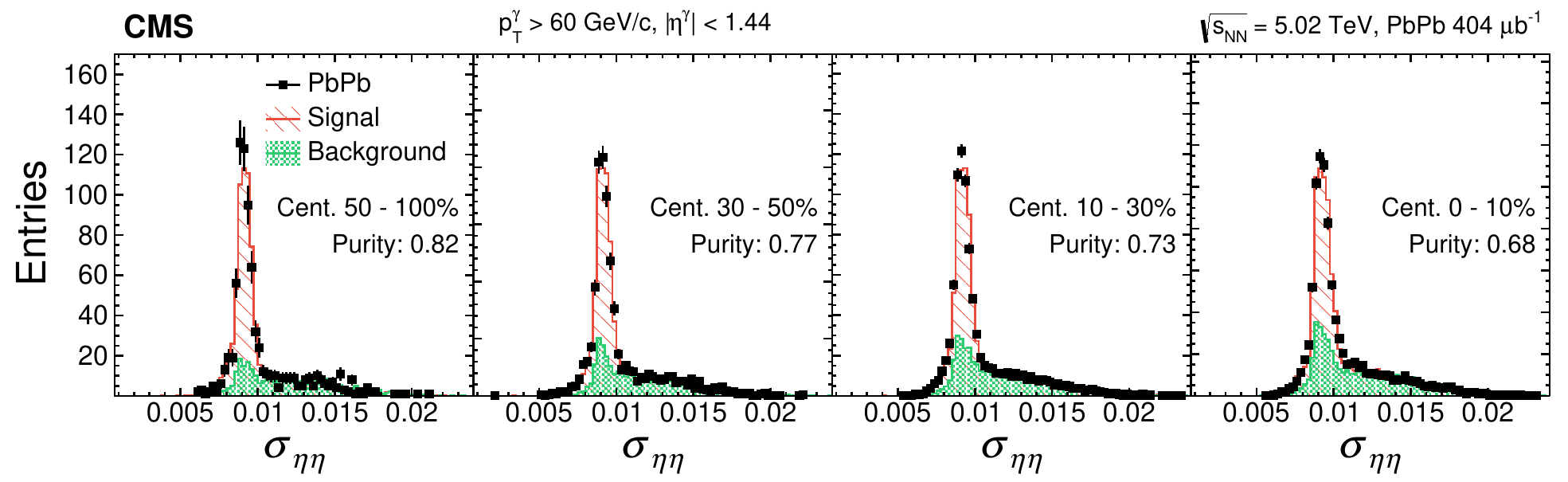}
    \caption{The centrality dependence of the shower shape variable
    $\sigma_{\eta\eta}$ for photons with $\ptg > 60\GeVc$. The black points
    show the \PbPb experimental results, the red histograms are the signal
    templates from \pythiahydjet simulations, and the green histograms are the
    background templates obtained from a nonisolated sideband region in data. }
    \label{fig:PbPb-purity}
\end{figure*}

The background contribution from pairs of decay photons and jets is subtracted
with a procedure based on collision data, using a two-component template fit of
the electromagnetic shower shape variable $\sigma_{\eta\eta}$, which is defined
as a modified second moment of the ECAL energy cluster distribution around its
mean $\eta$ position~\cite{Khachatryan:2010fm,AWES1992130}:

\ifthenelse{\boolean{cms@external}}{
\begin{multline}
\label{sieieFormula}
\sigma_{\eta\eta}^2 = \frac{\sum_i^{5{\times}5}w_i(\eta_i-\eta_{5{\times}5})^2}{\sum_i^{5{\times}5} w_i}, \\
 w_i = \max\left(0, 4.7 + \ln \frac{E_i}{E_{5{\times}5}}\right),
\end{multline}
}{
\begin{equation}
\label{sieieFormula}
\sigma_{\eta\eta}^2 = \frac{\sum_i^{5{\times}5}w_i(\eta_i-\eta_{5{\times}5})^2}{\sum_i^{5{\times}5} w_i}, \qquad w_i = \max\left(0, 4.7 + \ln \frac{E_i}{E_{5{\times}5}}\right),
\end{equation}
}
where $E_i$ and $\eta_i$ are the energy deposit and $\eta$ of the $i$th ECAL
crystal within a $5{\times}5$ crystal array centered around the electromagnetic
cluster, and $E_{5{\times}5}$ and $\eta_{5{\times}5}$ are the total energy and
mean $\eta$ of the $5{\times}5$ crystal matrix, respectively. The shape of the
signal distribution is obtained from \pythiahydjet simulations of isolated
prompt photon+jet processes, while the background templates are obtained from
a nonisolated sideband region in data, $10 < \sumIso < 20\GeVc$. The purity of
the photon sample (fraction of prompt photons within the remaining collection
of candidates) is determined from the fit. Examples of the template fits are
shown in Fig.~\ref{fig:PbPb-purity} for the lowest \ptg photons and the four
centrality intervals used in this analysis. The purity decreases in more
central collisions, reflecting an increase in the backgrounds.

The yields and kinematic characteristics of the background arising from pairs
of decay photons and jets are estimated by analyzing events with a larger
photon shower width ($0.011<\sigma_{\eta\eta}<0.017$), which are dominated by
decay photons. The background contribution fraction is then subtracted from the
yield for the signal events, which have a smaller photon shower width
($\sigma_{\eta\eta}<0.01$), according to the purity obtained from the template
fits.

The detector response for low-\pt jets can exhibit significant nonlinearity and
biases because of the background subtraction procedure of the current jet
algorithm, as well as the high magnetic field of the CMS detector. This is
neither well-modeled nor well-understood. Hence, the distributions are not
unfolded for the detector resolution, but the approach instead is to smear,
\ie, convolve with a Gaussian resolution adjustment term, the jet energy in \pp
events to match the JER in each of the \PbPb centrality classes in which the
comparison is made. This is done in every figure except
Fig.~\ref{fig:xjg-theory-pp}. The JER $\sigma(\ptGen)$ is defined as the
Gaussian standard deviation of the $\ptReco/\ptGen$ ratio, where $\ptReco$ is
the UE-subtracted, detector-level jet \pt, and $\ptGen$ is the generator-level
jet \pt without any contributions from a \PbPb UE. For \PbPb (\pp) collisions,
the JER is calculated from \pythiahydjet (\PYTHIA) events that are propagated
through the \GEANTfour~\cite{geant4} package. The UE produced by \HYDJET with
\GEANTfour simulation has been compared to data by observing the energy
collected inside randomly oriented cones with the same radius as the distance
parameter of the jet algorithm. The MC simulation is found to be in good
agreement with the experimental results. The JER is parametrized using the
expression

\begin{equation}
\sigma\left(\ptGen\right) = \sqrt{C^{2}+{\frac{S^{2}}{\ptGen}}+{\frac{N^{2}}{\left( \ptGen \right)^2}}} .
\label{eqn:jer}
\end{equation}

The stochastic term $S$ describes the \pt dependence of the jet energy
resolution, the constant term $C$ represents the high-\pt limit of the
resolution, and the noise term $N$ reflects the effect of UE fluctuations on
the energy resolution. All parameters for $\sigma(\ptGen)$ are determined using
\PYTHIA and \pythiahydjet samples with their numerical values provided in
Table~\ref{tab:resolution}. Following the smearing to 0--30\% PbPb data, the
energy resolutions of jets with $\ptj=30 (60)$\GeVc measured in pp data changes
from 18\%(14\%) to 35\%(22\%) respectively. Compared to the JER, the jet $\phi$
resolution has a negligible effect.

\begin{table*}[ht]
\centering
\topcaption{Jet resolution parameters for \pp and \PbPb collisions. A global
uncertainty of 7\% is assigned to the smearing parameters, evaluated as
described in text.}
\label{tab:resolution}
\newcolumntype{z}{D{,}{\text{--}}{2.3}}
\begin{tabular}{c z c c c }
& \multicolumn{1}{c}{Centrality [\%]} & $C$                   & $S$ [$(\GeVcns{})^{1/2}$] & $N$ [\GeVcns{}] \\
\hline
pp &  \multicolumn{1}{c}{\ensuremath{\text{---}}}  & 0.06   & 0.95 & 0           \\[\cmsTabSkip]
\multirow{5}{*}{PbPb}
& 0, 30                             & \multirow{2}{*}{0.06} & \multirow{2}{*}{1.24} & 6.83       \\
& 30, 100                            &                       &                       & 0           \\[\cmsTabSkip]
& 0, 10                             & \multirow{4}{*}{0.06} & \multirow{4}{*}{1.24} & 8.42       \\
& 10, 30                             &                       &                       & 5.54       \\
& 30, 50                             &                       &                       & 2.37       \\
& 50, 100                            &                       &                       & 0           \\
\hline
\end{tabular}
\end{table*}

\subsection{Systematic uncertainties}
\label{sec:systematics}

Systematic uncertainties are estimated separately for the \pp and \PbPb
analyses. The uncertainties are determined for each centrality and \ptg
interval using similar procedures as described in
Ref.~\cite{Chatrchyan:2012gt}. Seven sources of uncertainty are considered:
photon purity, isolation definition, photon energy scale, electron
contamination, photon efficiency, JER, and jet energy scale (JES). The total
systematic uncertainties are calculated by summing in quadrature the
uncertainties from all sources.

The uncertainty on the photon purity estimate is evaluated by varying the
nonisolated sideband regions used to obtain the background template. The
maximum deviation from the nominal values is $\pm$10\%\,($\pm$6\%) for central
(peripheral) \PbPb collisions, and $\pm$5\% in \pp collisions. The varied
purity values are then used to perform the background subtraction, and the
maximum difference from the nominal results is quoted as the uncertainty. The
uncertainty due to the isolated photon definition is determined by comparing
the photon+jet observables when using generator-level and detector-level
definitions of the isolation variables. The photon energy scale uncertainty is
based on the residual data-to-simulation photon energy scale difference after
applying the photon energy corrections, amounting to about 1\%, independent of
\ptg and event centrality. The uncertainty due to electron contamination is
evaluated by repeating the analysis without applying electron rejection, and
scaling the difference in the final observables to the residual electron
contamination after applying electron rejection. The electron rejection
efficiency is determined to be 66\% from MC studies. The uncertainty on the
photon efficiency correction is determined by varying the selection criteria
for matching reconstructed photons with generator-level photons. The
uncertainty on the JER has two sources. The first source is the difference
between the JER in data and simulation, which is around 15\% for all
centralities in both \pp and \PbPb collisions. The associated systematic
uncertainty is evaluated by propagating the effects of having a JER that
differs by 15\% relative to the nominal value. The second source (7\%) accounts
for the uncertainty in the resolution and the modeling of the JER
distributions, and was obtained by considering the differences between the
extracted JER in each \ptGen bin and the parametrization using
Eq.~\ref{eqn:jer}, and determining the value at one standard deviation of that
distribution, assuming that the differences are normally distributed.

Finally, the JES uncertainty arises from three contributions that are added in
quadrature for the final value. Two are common to both the \pp and \PbPb
samples: the residual deviation from unity in simulation (\ie, the closure) of
the JES after applying all jet energy corrections (2\%) and the difference
between data and simulation (2\%). These two effects are independent of
centrality and together amount to 2.8\%. The closure of the JES depends on the
flavor of the fragmenting parton: simulations show that the energy scale of
quark jets is consistently higher than that of gluon jets. For \pp collisions,
the fragmentation dependence of the JES has been studied and is accounted for
in the uncertainty from the difference between data and simulation. However, in
\PbPb collisions, the ratio of quarks and gluons can be different from \pp data
because of expected differences in centrality-dependent quenching of jets
initiated by quarks or gluons. The subtraction of the UE in \PbPb collisions
results in the JES having a larger dependence on the fragmentation pattern than
found for \pp collisions, since one can only distinguish between soft particles
from the jet fragmentation and the underlying event on average. Hence, an
additional uncertainty, evaluated using collision data and simulation, is
applied in \PbPb collisions to account for these fragmentation effects on the
JES arising from the subtraction algorithm, underlying event, and quenching.
The photon-tagged jet fragmentation function in \PbPb data is constructed and
fit by a two-component model of the jet fragmentation functions for quark and
gluon jets that were obtained from MC simulations. For $\ptg > 60\GeVc$, the
results show that the fraction of jets originating from gluon fragmentation in
data can be constrained to between 0\% and approximately 26\%, which
corresponds to the fraction found in \pythiahydjet MC samples. Hence, in this
kinematic region, the difference between the JES for a pure quark jet sample
and the inclusive sample is used in the uncertainty estimation. For $40 < \ptg
< 60\GeVc$, where the results of the template fit are inconclusive because of
the large statistical uncertainties, the full difference in the JES between
having 0\% and 100\% gluon jet fraction is used. This difference is
approximately 2--5\%\,(1.5--2.5\%) in central (peripheral) collisions. The
final systematic uncertainty associated with the unknown quark-gluon ratio in
data is taken as the maximum deviation from varying the JES up and down
according to the quark-gluon ratio constraints mentioned above for each \ptg
interval.

A summary of the systematic uncertainties for \rjg, \avexjg, and \dphijg in
\PbPb collisions is shown in Tables~\ref{table:sys_unc_totals} and
\ref{table:sys_unc_dphi}, averaged over multiple \ptg and/or event centrality
intervals. The dominant sources of uncertainties in both \pp and \PbPb
collisions are from JES and photon purity estimation. The systematic
uncertainties for \PbPb and \pp collisions are considered uncorrelated.

\begin{table*}[tb]
\centering
\topcaption{\label{table:sys_unc_totals} Summary of the relative systematic
uncertainties (in \%) for $\ptg > 40$\GeVc.}
\begin{tabular}{lcccccc}
                       & \multicolumn{2}{c}{\pp} & \multicolumn{4}{c}{\PbPb}                                                         \\
\hline
Source of systematic   & \multicolumn{2}{c}{}    & \multicolumn{2}{c}{0--30\% Centrality} & \multicolumn{2}{c}{30--100\% Centrality} \\ \cline{4-7}
uncertainty [\%]       & \avexjg    & \rjg       & \avexjg & \rjg     & \avexjg & \rjg \\
\hline
Photon energy scale    & ${<}0.5$   & ${<}0.5$   & 0.7     & ${<}0.5$ & ${<}0.5$ & 0.5 \\
Photon isolation       & 0.8        & 0.9        & 0.8     & 1.0      & 0.8      & 0.7 \\
Photon purity          & ${<}0.5$   & 0.5        & 3.1     & 3.5      & 2.0      & 2.2 \\
Photon efficiency      & ${<}0.5$   & ${<}0.5$   & ${<}0.5$& ${<}0.5$ & ${<}0.5$ & ${<}0.5$ \\
Electron contamination & ${<}0.5$   & ${<}0.5$   & 0.5     & 0.9      & ${<}0.5$ & 0.9 \\
Jet energy scale       & 1.9        & 1.8        & 2.8     & 7.3      & 2.8      & 5.1 \\
Jet energy resolution  & 0.9        & 1.1        & 2.3     & 3.6      & 1.0      & 1.5 \\
\hline
\end{tabular}
\end{table*}

\begin{table*}[bt]
\centering
\topcaption{\label{table:sys_unc_dphi} Summary of the absolute systematic
uncertainties on $({1}/{N_{{\mathrm{j}}\gamma}})
({\rd{}N}/{\rd{}\phi_{\mathrm{j}\gamma}})$ for $\ptg
> 40$\GeVc, averaged over the \dphijg distributions.
}
\begin{tabular}{lccc}
                       & \pp                      & \multicolumn{2}{c}{\PbPb}                                                   \\
\hline
Source of systematic   &                          & \multirow{2}{*}{0--30\% Centrality} & \multirow{2}{*}{30--100\% Centrality} \\
uncertainty            &                          &                                     &                                       \\ \hline
Photon energy scale    & ${<}0.01 \times 10^{-2}$ & $2.12 \times 10^{-2}$               & $0.08 \times 10^{-2}$                 \\
Photon isolation       & $0.27 \times 10^{-2}$    & $0.26 \times 10^{-2}$               & $0.16 \times 10^{-2}$                 \\
Photon purity          & $0.13 \times 10^{-2}$    & $0.78 \times 10^{-2}$               & $0.61 \times 10^{-2}$                 \\
Photon efficiency      & ${<}0.01 \times 10^{-2}$ & $0.09 \times 10^{-2}$               & $0.03 \times 10^{-2}$                 \\
Electron contamination & $0.05 \times 10^{-2}$    & $0.19 \times 10^{-2}$               & $0.14 \times 10^{-2}$                 \\
Jet energy scale       & $0.23 \times 10^{-2}$    & $1.63 \times 10^{-2}$               & $0.86 \times 10^{-2}$                 \\
Jet energy resolution  & $0.31 \times 10^{-2}$    & $0.46 \times 10^{-2}$               & $0.48 \times 10^{-2}$                 \\
\hline
\end{tabular}
\end{table*}

\section{Results and discussion}
\label{sec:results}

\subsection{Photon+jet azimuthal correlation}

\begin{figure*}[ht]
\centering
\includegraphics[width=0.98\textwidth]{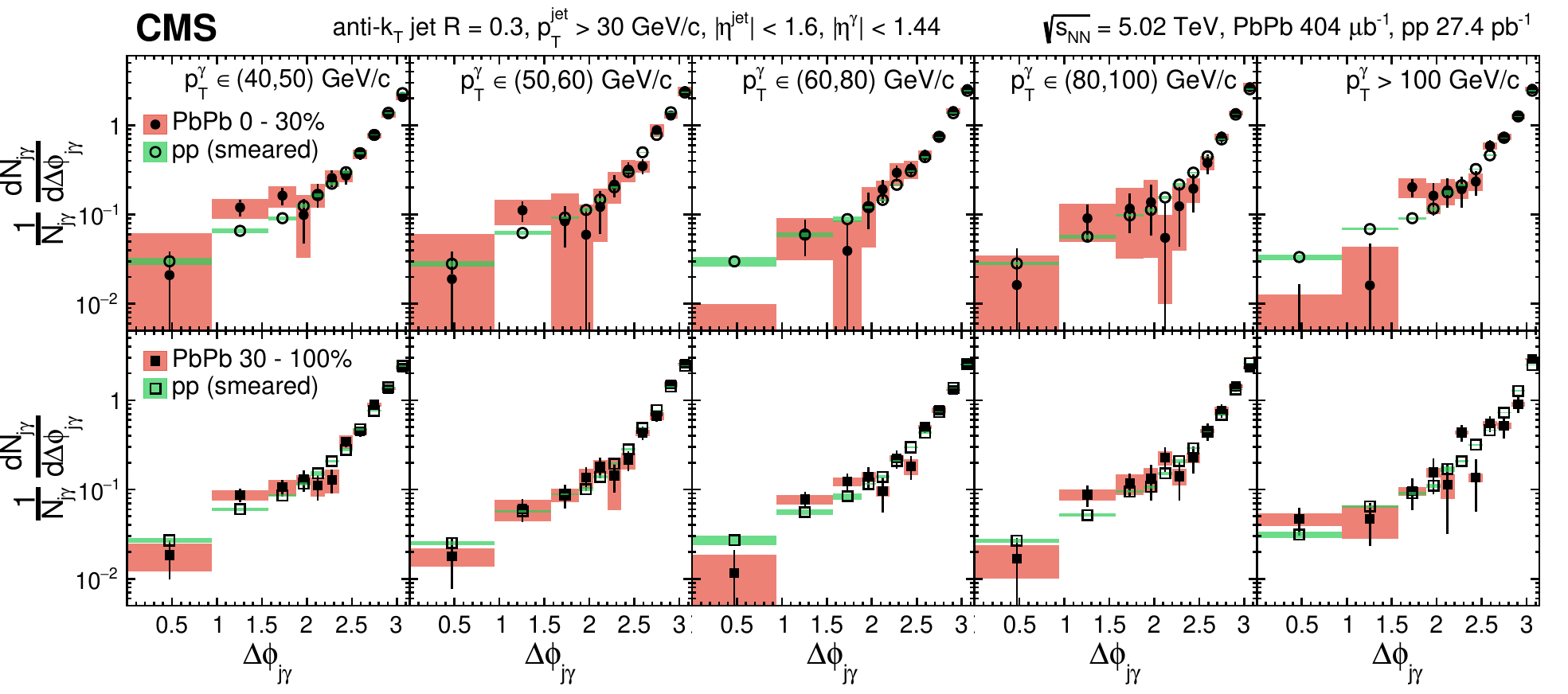}
\caption{\label{fig:dphi_log}
The azimuthal correlation of photons and jets in five \ptg intervals for
0--30\% centrality (top, full circles) and 30--100\% centrality (bottom, full
squares) \PbPb collisions. The smeared \pp data (open symbols) are included for
comparison. The vertical lines (bands) through the points represent statistical
(systematic) uncertainties.
}
\end{figure*}

Possible modification of the back-to-back photon and recoiling jet alignment by
the medium can be studied by comparing the relative azimuthal angle (\dphijg)
distributions in \pp and \PbPb collisions~\cite{Chatrchyan:2011sx,Aad:2010bu}.
The distributions are normalized by the number of photon+jet pairs. The shape
of the \dphijg distribution in \pp and \PbPb collisions is studied in intervals
of leading photon \pt and two event centrality classes, as shown in
Fig.~\ref{fig:dphi_log}. The exponentially falling region ($\dphijg > 2\pi/3$)
is fit to a normalized exponential function, as in
Ref.~\cite{Chatrchyan:2012gt}, and the values of the exponents in \PbPb and \pp
collisions from the fits are compared. Within the quoted statistical and
systematic uncertainties, the \PbPb results with different photon \pt and event
centrality selections are consistent with the corresponding smeared \pp
reference data, \ie, no broadening of the distributions is observed.

\subsection{Photon+jet transverse momentum imbalance}

\begin{figure*}[ht]
\centering
\includegraphics[width=0.98\textwidth]{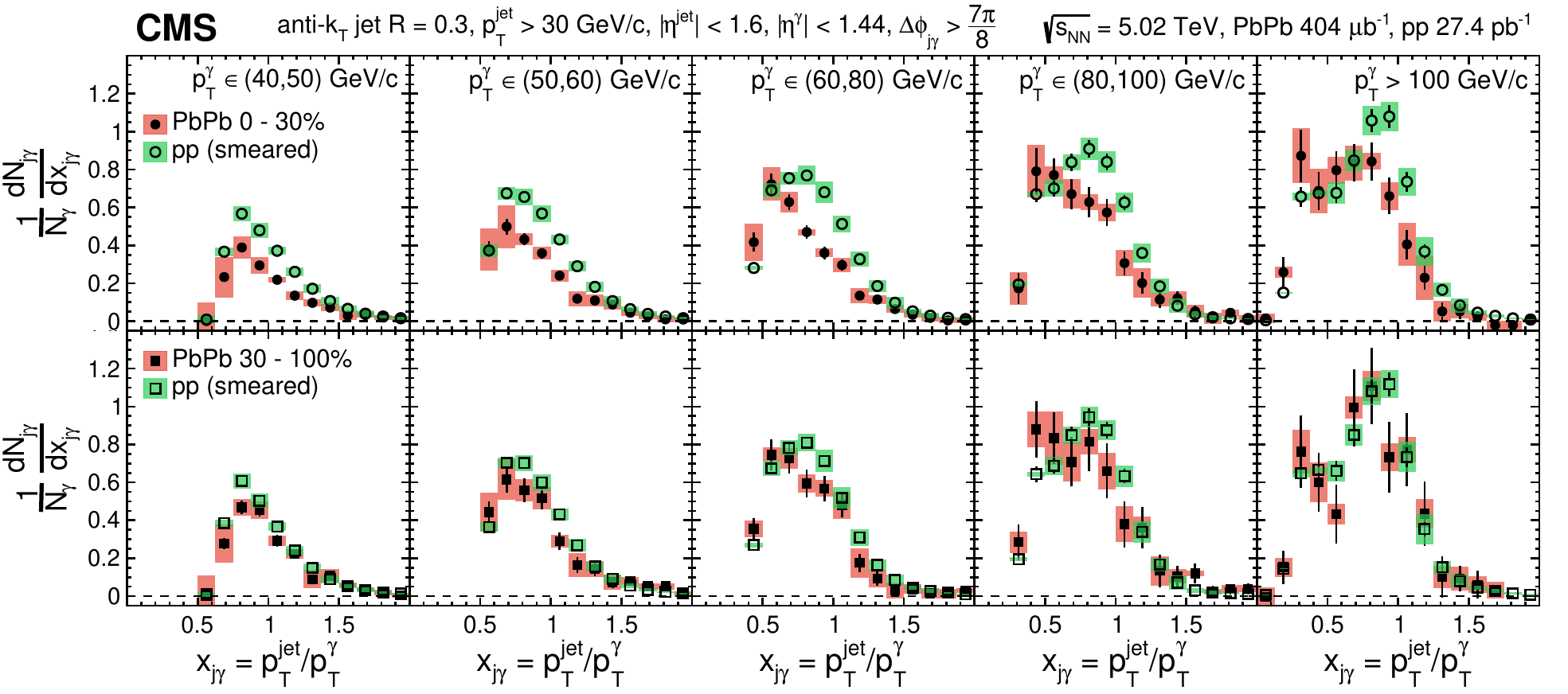}
\caption{\label{fig:xjg}
Distribution of $\xjg = \ptj/ \ptg$ in five \ptg intervals for 0--30\%
centrality (top, full circles) and 30--100\% centrality (bottom, full squares)
\PbPb collisions.
The smeared \pp data (open symbols) are included for comparison. The vertical
lines (bands) through the points represent statistical (systematic)
uncertainties.
}
\end{figure*}

The asymmetry ratio $\xjg = \ptj/ \ptg$ is used to quantify the photon+jet \pt
imbalance due to in-medium parton energy loss. In addition to the photon and
jet selections used in the $\dphijg$ study, a $\dphijg > (7\pi)/8$ selection is
applied to select back-to-back photon+jet topologies, suppressing the
contributions from background jets as well as photon-multijet events.
Figure~\ref{fig:xjg} shows the \xjg distributions for different centrality and
\ptg regions in \pp and \PbPb collisions, normalized by the number of photons.
In 0--30\% centrality \PbPb collisions, significant modifications (lower mean
and smaller integral values) of the \xjg spectra with respect to the smeared
\pp reference data are observed, while the modifications are smaller in the
30--100\% centrality \PbPb collisions.

\begin{figure}[ht]
\centering
\includegraphics[width=\cmsFigWidth]{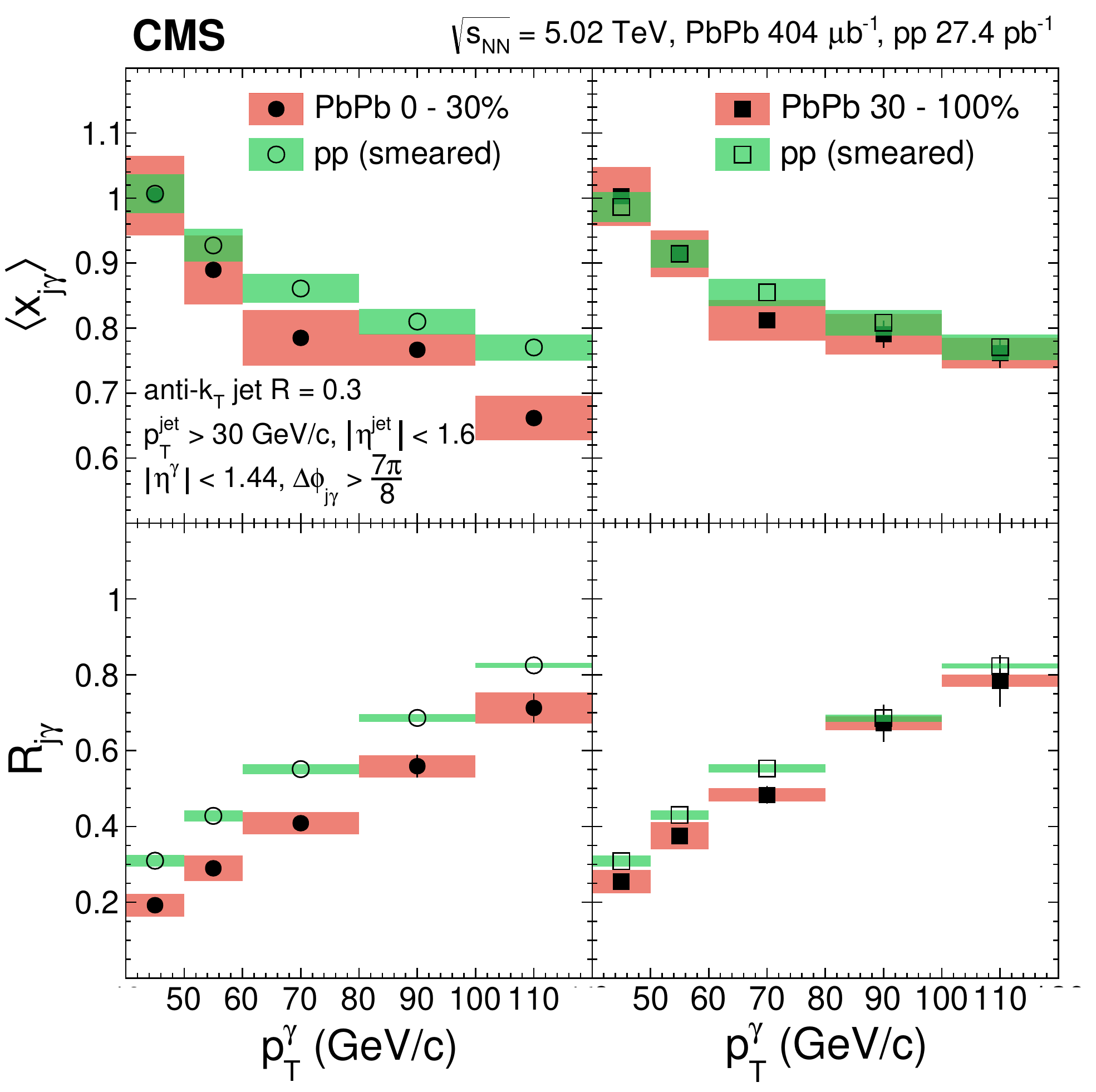}
\caption{\label{fig:xjg_mean_rjg_ptBinAll}
The \avexjg values (top) and \rjg, the number of
associated jets per photon (bottom), in 0--30\% centrality (left, full circles)
and 30--100\% centrality (right, full squares) \PbPb collisions. The smeared
\pp data (open symbols) are added for comparison. The vertical lines (bands)
through the points represent statistical (systematic) uncertainties.
}
\end{figure}

The mean values, \avexjg (in effect, a truncated mean because of the \ptj
threshold), of the \xjg distributions are shown as a function of \ptg in
Fig.~\ref{fig:xjg_mean_rjg_ptBinAll} (top). The \avexjg values in \PbPb and
smeared \pp collisions are consistent with each other within the quoted
uncertainties over the whole \ptg interval probed in 30--100\% centrality \PbPb
collisions and in the region $\ptg < 60$\GeVc for 0--30\% centrality \PbPb
collisions. At higher \ptg in the more central \PbPb events, the \avexjg value
is lower than in \pp data.

With a jet \pt threshold of 30\GeVc, the \avexjg values observed for the
selected photon+jet pairs likely underestimates the actual imbalance.
Photon+jet pairs for which the momentum of the associated jets falls below the
jet \pt threshold do not contribute to the \avexjg value. To assess how the
``missing'' jets might affect the \avexjg results, the average number of
associated jets per photon passing the analysis selections, \rjg, is shown in
Fig.~\ref{fig:xjg_mean_rjg_ptBinAll} (bottom). In the 0--30\% most central
\PbPb collisions, the value of \rjg is found to be lower than in the smeared
\pp data in all leading photon \pt intervals. The absolute difference is
approximately constant as a function of \ptg, but the relative difference is
larger at lower \ptg, since the \rjg in \pp collisions is itself lower in that
region.

\subsection{Jet yield ratio}

\begin{figure*}[ht]
\centering
\includegraphics[width=0.98\textwidth]{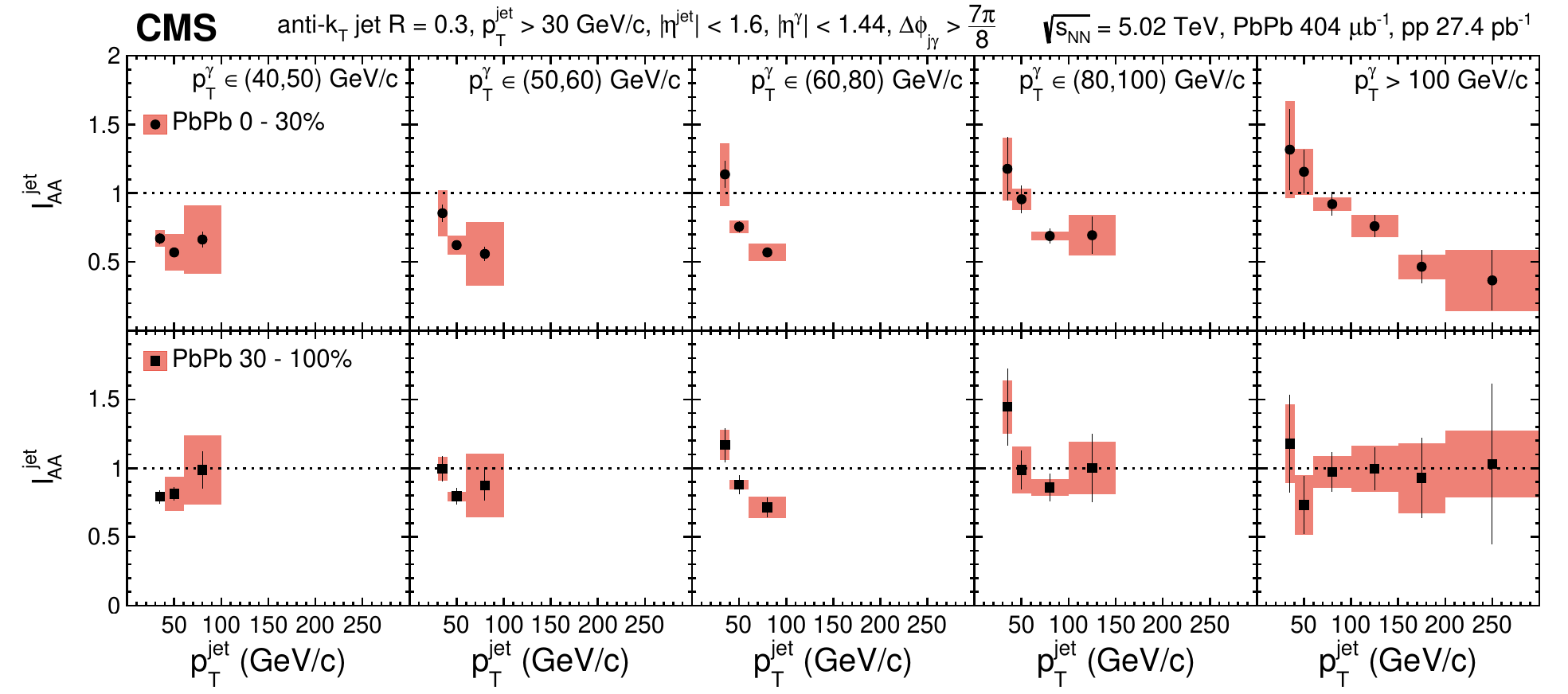}
\caption{\label{fig:iaa}
The \iaa \vs \ptj for 0--30\% centrality (top) and 30--100\% centrality
(bottom) \PbPb collisions. The vertical lines (bands) through the points
represent statistical (systematic) uncertainties.
}
\end{figure*}

Figure~\ref{fig:iaa} shows, as a function of \ptj for several \ptg intervals
and two \PbPb event centrality intervals, the ratio of the associated jet
yields in \PbPb and smeared \pp events, \iaa:

\begin{equation}
\label{eq:jetiaa-eqn}
\mathrm{I_{AA}^{jet}} = \left( \frac{1}{N^{\gamma}_{\text{\PbPb}}} \frac{\rd N^{\text{jet}}_{\PbPb}}{\rd \pt^{\text{jet}}} \right) \Bigg/ \left( \frac{1}{N^{\gamma}_{\pp}} \frac{\rd N^{\text {jet}}_{\pp}}{\rd\pt^{\text{jet}}} \right).
\end{equation}

This variable reflects the modification of the associated jet \pt spectra by
the medium. In 30--100\% \PbPb events, the \iaa values are slightly suppressed
for photon candidates with $\ptg < 80$ \GeVc, and consistent with unity for
photon candidates with $\ptg > 80$ \GeVc. For 0--30\% centrality \PbPb events,
a suppression of approximately a factor of 2 is observed at low \ptg. As the
\ptg increases, the larger phase space allows quenched jets to remain above the
kinematic selections, which translates to a slight excess of quenched jets
appearing at low \ptj. This is seen in the top row, where \iaa for low \ptj
increases with \ptg while the \iaa at large \ptj stays roughly constant.

\subsection{Centrality dependence}

\begin{figure*}[ht]
\centering
\includegraphics[width=0.98\textwidth]{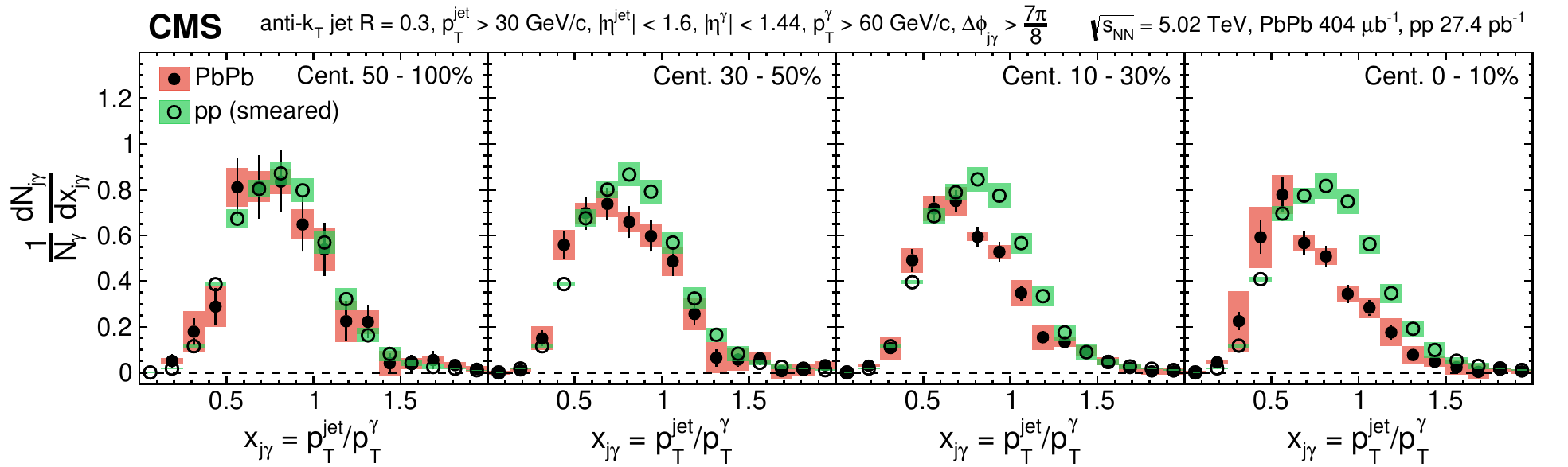}
\caption{\label{fig:xjg_cent}
The centrality dependence of \xjg of photon+jet pairs normalized by the number
of photons for \PbPb (full markers) and smeared \pp (open markers) data. The
vertical lines (bands) through the points represent statistical (systematic)
uncertainties.
}
\end{figure*}

The centrality dependence in \PbPb collisions of \xjg spectra for
$\ptg>60\GeVc$ is shown in Fig.~\ref{fig:xjg_cent}. In the most peripheral
collisions (50--100\% centrality), the \xjg distribution agrees with the
smeared \pp reference data. As collisions become more central, the \PbPb
distributions shift towards lower \xjg and the integrals of the \xjg spectra
become smaller. This is consistent with the expectation that a larger amount of
parton \pt is transported out of the jet cone as a consequence of the larger
average path length that the parton needs to travel through in more central
\PbPb collisions~\cite{Aad:2012vca,Khachatryan:2015lha}.

\begin{figure}[htb]
\centering
\includegraphics[width=\cmsFigWidth]{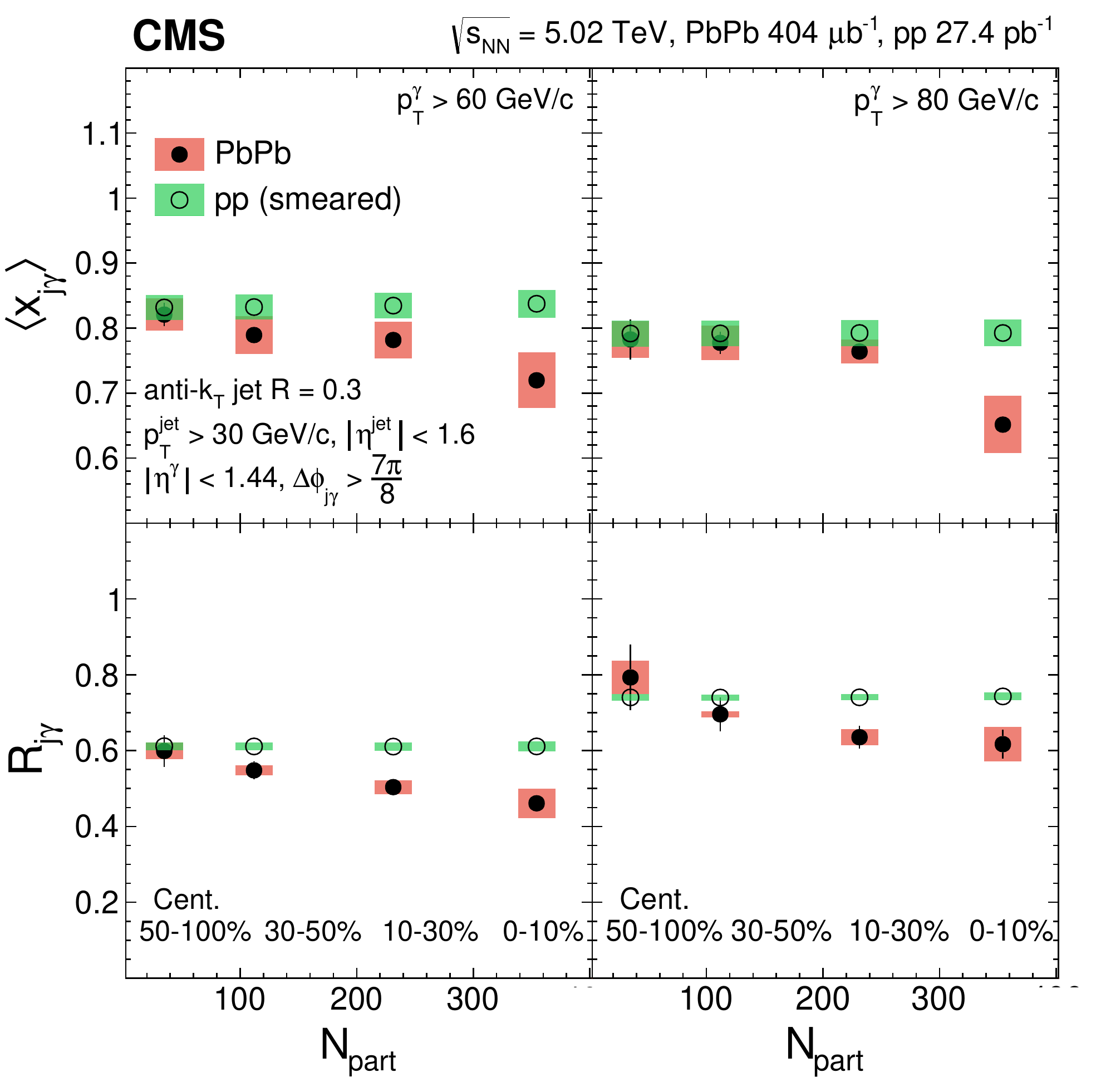}
\caption{\label{fig:xjg_mean_rjg_centBinAll}
The \avexjg (top) and \rjg (bottom) as a function of \npart for
$\ptg>60$\GeVc (left) and $\ptg>80$\GeVc (right). The \PbPb results (full
markers) are compared to \pp results (open markers) smeared by the relative jet
energy resolution corresponding to each centrality interval. The vertical lines
(bands) through the points represent statistical (systematic) uncertainties.
}
\end{figure}

Figure~\ref{fig:xjg_mean_rjg_centBinAll} shows \avexjg and \rjg in \pp and
\PbPb collisions as a function of event centrality, quantified by \npart, which
is the mean number of participating nucleons within a given centrality
interval. The \npart values are estimated from a MC Glauber
model~\cite{Miller:2007ri, Khachatryan:2016odn}. In central collisions,
a suppression of both \avexjg and \rjg is observed in comparison to the smeared
\pp reference data, consistent with significant in-medium energy loss of the
associated jets.

\subsection{Comparison to theoretical models}

\begin{figure*}[htb]
\centering
\includegraphics[width=0.98\textwidth]{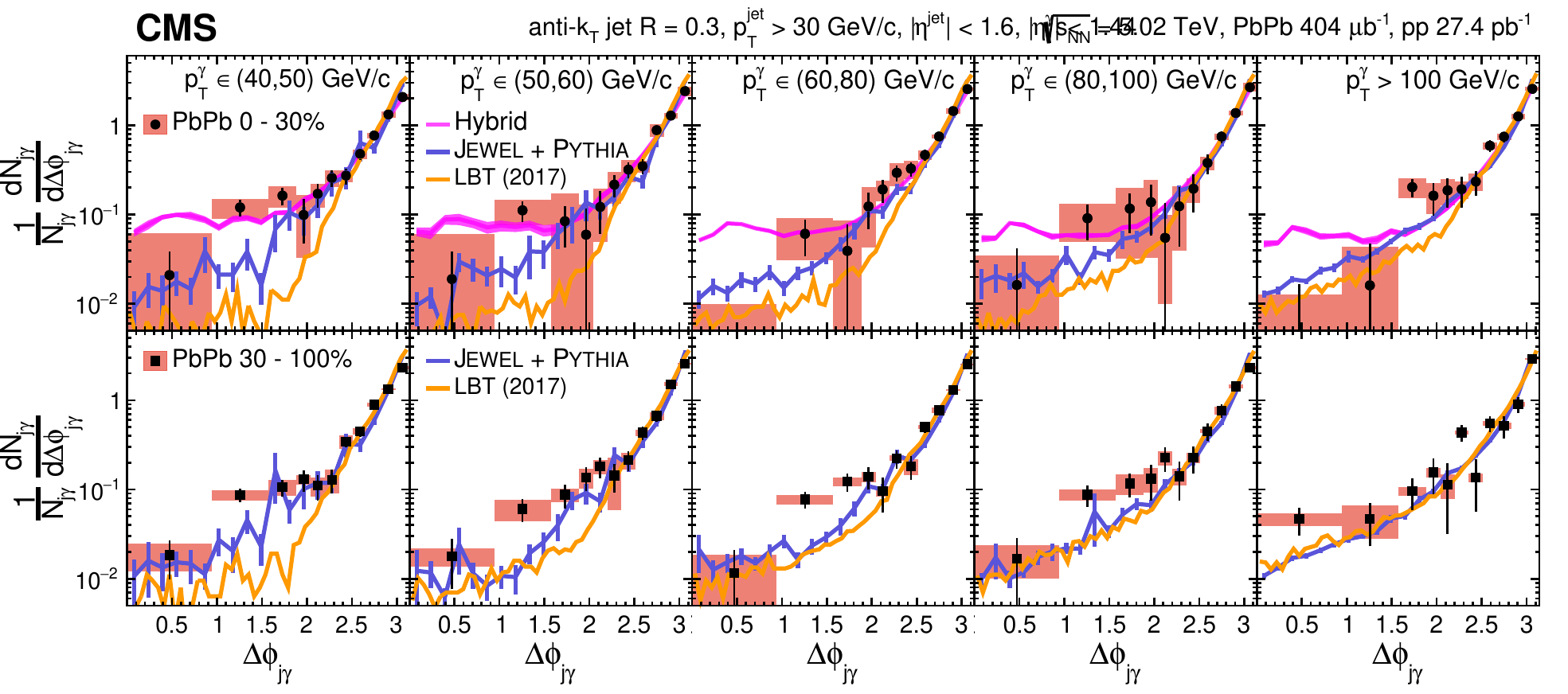}
\caption{\label{fig:dphi-theory-PbPb}
The azimuthal correlation of photons and jets in five \ptg intervals for
0--30\% centrality (top, full circles) and 30--100\% centrality (bottom, full
squares) \PbPb collisions. The data points shown are identical to those in
Fig.~\ref{fig:dphi_log}. Theoretical calculations from
\jewel~\cite{jewel:2013,jewel:2016}, LBT~\cite{Wang:2013cia}, and hybrid
model~\cite{hybrid:2014,hybrid:2015} are included for comparison.
}
\end{figure*}

\begin{figure*}[htb]
\centering
\includegraphics[width=0.98\textwidth]{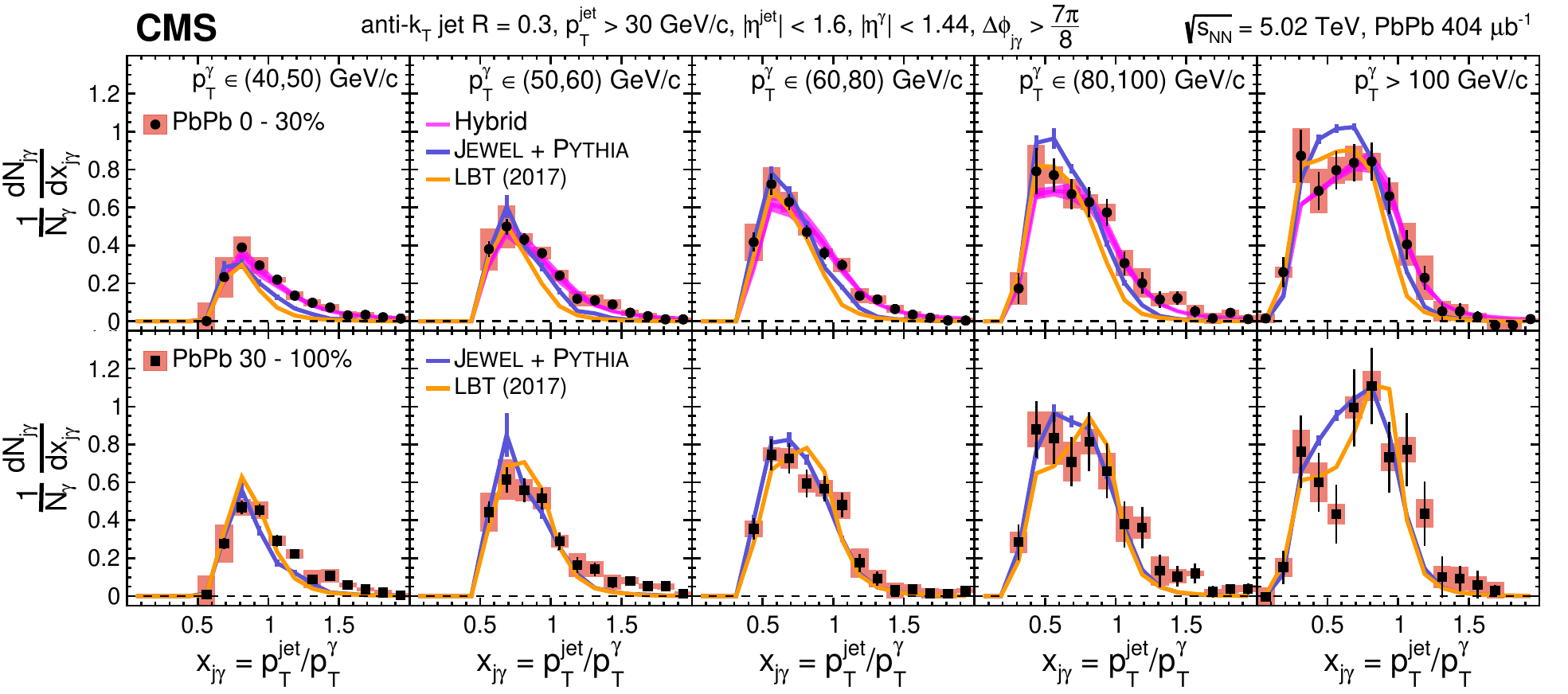}
\caption{\label{fig:xjg-theory-PbPb}
The \xjg distributions in five \ptg intervals for 0--30\% centrality (top, full
circles) and 30--100\% centrality (bottom, full squares) \PbPb collisions. The
data points shown are identical to those in Fig.~\ref{fig:xjg}. Theoretical
calculations from \jewel~\cite{jewel:2013,jewel:2016}, LBT~\cite{Wang:2013cia},
and hybrid model~\cite{hybrid:2014,hybrid:2015} are included for comparison.
}
\end{figure*}

\begin{figure*}[htb]
\centering
\includegraphics[width=0.98\textwidth]{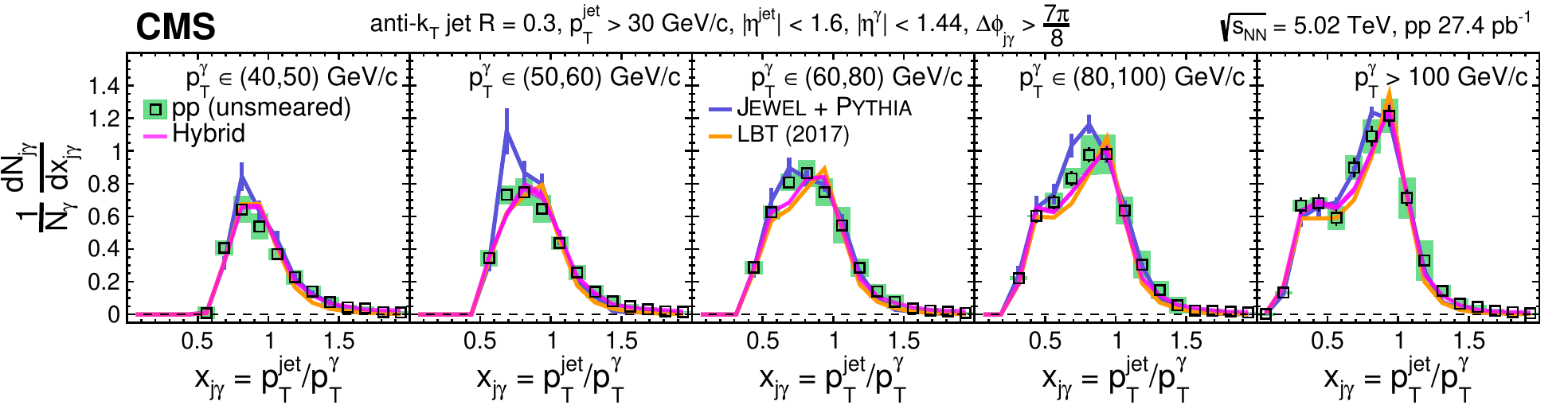}
\caption{\label{fig:xjg-theory-pp}
The \xjg distributions in five \ptg intervals for unsmeared \pp data (full
squares). The \xjg distributions in \pp collisions assumed by the
\jewel~\cite{jewel:2013,jewel:2016}, LBT~\cite{Wang:2013cia}, and hybrid
models~\cite{hybrid:2014,hybrid:2015} discussed in this Letter are also shown
for comparison.
}
\end{figure*}

The results for \PbPb collisions presented in Fig.~\ref{fig:dphi_log} for
\dphijg and Fig.~\ref{fig:xjg} for \xjg are compared with several theoretical
calculations with different approaches to modeling the jet energy loss in
Figs.~\ref{fig:dphi-theory-PbPb} and \ref{fig:xjg-theory-PbPb}, respectively.
The \xjg distributions assumed by the different model calculations in \pp
collisions are compared to the unsmeared \pp data in
Fig.~\ref{fig:xjg-theory-pp}. The \jewel model is a dynamical, perturbative
framework for jet quenching, which has been extended to simulate boson-jet
events~\cite{jewel:2013,jewel:2016}. The LBT 2017 model~\cite{Wang:2013cia}
uses a linearized Boltzmann transport model for jet propagation through the
medium, including the recoiled medium partons in the reconstruction of the
partonic jets. The hybrid model~\cite{hybrid:2014,hybrid:2015} combines
a perturbative description of the weakly coupled physics of jet production and
evolution with a gauge/gravity duality description of the strongly coupled
dynamics of the medium, and of the soft exchanges between the jet and the
medium. The calculations from the \jewel and hybrid models have been smeared to
the corresponding JER in \pp or \PbPb collisions.

Predictions from the \jewel and hybrid models have previously shown reasonable
agreement with measurements of inclusive jet nuclear modification
factors~\cite{Zapp:2012ak,hybrid:2015}. For the results reported in this
Letter, all models describe well the \pp results. They also capture the general
features of the 0--30\% \PbPb data, although the hybrid model appears to better
describe the \xjg results. As shown in Fig.~\ref{fig:xjg-theory-PbPb}, the
\jewel and LBT models appear to underestimate the \xjg spectra in the high \xjg
region ($\xjg>0.9$) for central \PbPb collisions, which suggests that the
amount of energy transported out of the jet cone is larger in these models than
in data. A similar effect is also hinted at in the 30--100\% \PbPb data, which
can be attributed to the fact that those distributions are dominated by events
in the 30--50\% centrality interval, where energy loss effects are still
significant. The models are also consistent with data in that none of them show
a broadening of the observed \dphijg distributions in \PbPb compared to \pp
collisions in the photon and jet kinematic ranges presented, despite their
implementing contributions from partonic collisions.

\section{Summary}
\label{sec:conclusion}

Correlations of isolated photons with transverse momentum $\ptg>40$\GeVc and
pseudorapidity $\abs{\etag}<1.44$ and associated jets with $\ptj>30$\GeVc and
$\abs{\etaj}<1.6$, have been studied for the first time in \pp and \PbPb
collisions at $\sqrtsNN = 5.02\TeV$, using a large data sample collected by the
CMS experiment. No significant azimuthal angular broadening between photons and
the associated jets is observed in \PbPb data as compared to \pp data, for all
event centralities and multiple photon \pt intervals. The $\xjg = \ptj/\ptg$
and the average number of associated jets per photon, \rjg, are studied in
different leading photon \pt and \PbPb collision centrality intervals. For all
$\ptg>60$\GeVc intervals, the \avexjg and \rjg values in the 0--30\% most
central \PbPb collisions are found to be lower than those in the corresponding
\pp reference data, indicating that a larger fraction of jets lose energy and
thus fall below 30\GeVc in \PbPb collisions. The differences between the \pp
and \PbPb results increase as collisions become more central. A shift of the
jet spectra towards lower \ptj is observed when comparing the yields of
associated jets in the 0--30\% most central \PbPb collisions to those in \pp
collisions. These new results are qualitatively similar to those reported at
$\sqrtsNN = 2.76$\TeV and to calculations from various theoretical models. The
better statistical precision of the new higher energy data provides an
opportunity to test theoretical models against data over a wide kinematic range
in \ptg and \xjg, and for different event centralities, using a selection of
partons with defined flavor (quark/gluon) and initial kinematics.

\ifthenelse{\boolean{cms@external}}{\clearpage}{}
\begin{acknowledgments}
We congratulate our colleagues in the CERN accelerator departments for the excellent performance of the LHC and thank the technical and administrative staffs at CERN and at other CMS institutes for their contributions to the success of the CMS effort. In addition, we gratefully acknowledge the computing centers and personnel of the Worldwide LHC Computing Grid for delivering so effectively the computing infrastructure essential to our analyses. Finally, we acknowledge the enduring support for the construction and operation of the LHC and the CMS detector provided by the following funding agencies: BMWFW and FWF (Austria); FNRS and FWO (Belgium); CNPq, CAPES, FAPERJ, and FAPESP (Brazil); MES (Bulgaria); CERN; CAS, MoST, and NSFC (China); COLCIENCIAS (Colombia); MSES and CSF (Croatia); RPF (Cyprus); SENESCYT (Ecuador); MoER, ERC IUT, and ERDF (Estonia); Academy of Finland, MEC, and HIP (Finland); CEA and CNRS/IN2P3 (France); BMBF, DFG, and HGF (Germany); GSRT (Greece); OTKA and NIH (Hungary); DAE and DST (India); IPM (Iran); SFI (Ireland); INFN (Italy); MSIP and NRF (Republic of Korea); LAS (Lithuania); MOE and UM (Malaysia); BUAP, CINVESTAV, CONACYT, LNS, SEP, and UASLP-FAI (Mexico); MBIE (New Zealand); PAEC (Pakistan); MSHE and NSC (Poland); FCT (Portugal); JINR (Dubna); MON, RosAtom, RAS, RFBR and RAEP (Russia); MESTD (Serbia); SEIDI, CPAN, PCTI and FEDER (Spain); Swiss Funding Agencies (Switzerland); MST (Taipei); ThEPCenter, IPST, STAR, and NSTDA (Thailand); TUBITAK and TAEK (Turkey); NASU and SFFR (Ukraine); STFC (United Kingdom); DOE and NSF (USA).

\hyphenation{Rachada-pisek} Individuals have received support from the Marie-Curie program and the European Research Council and Horizon 2020 Grant, contract No. 675440 (European Union); the Leventis Foundation; the A. P. Sloan Foundation; the Alexander von Humboldt Foundation; the Belgian Federal Science Policy Office; the Fonds pour la Formation \`a la Recherche dans l'Industrie et dans l'Agriculture (FRIA-Belgium); the Agentschap voor Innovatie door Wetenschap en Technologie (IWT-Belgium); the Ministry of Education, Youth and Sports (MEYS) of the Czech Republic; the Council of Science and Industrial Research, India; the HOMING PLUS program of the Foundation for Polish Science, cofinanced from European Union, Regional Development Fund, the Mobility Plus program of the Ministry of Science and Higher Education, the National Science Center (Poland), contracts Harmonia 2014/14/M/ST2/00428, Opus 2014/13/B/ST2/02543, 2014/15/B/ST2/03998, and 2015/19/B/ST2/02861, Sonata-bis 2012/07/E/ST2/01406; the National Priorities Research Program by Qatar National Research Fund; the Programa Severo Ochoa del Principado de Asturias; the Thalis and Aristeia programs cofinanced by EU-ESF and the Greek NSRF; the Rachadapisek Sompot Fund for Postdoctoral Fellowship, Chulalongkorn University and the Chulalongkorn Academic into Its 2nd Century Project Advancement Project (Thailand); the Welch Foundation, contract C-1845; and the Weston Havens Foundation (USA).
\end{acknowledgments}

\bibliography{auto_generated}
\cleardoublepage \appendix\section{The CMS Collaboration \label{app:collab}}\begin{sloppypar}\hyphenpenalty=5000\widowpenalty=500\clubpenalty=5000\vskip\cmsinstskip
\textbf{Yerevan~Physics~Institute,~Yerevan,~Armenia}\\*[0pt]
A.M.~Sirunyan, A.~Tumasyan
\vskip\cmsinstskip
\textbf{Institut~f\"{u}r~Hochenergiephysik,~Wien,~Austria}\\*[0pt]
W.~Adam, F.~Ambrogi, E.~Asilar, T.~Bergauer, J.~Brandstetter, E.~Brondolin, M.~Dragicevic, J.~Er\"{o}, M.~Flechl, M.~Friedl, R.~Fr\"{u}hwirth\cmsAuthorMark{1}, V.M.~Ghete, J.~Grossmann, J.~Hrubec, M.~Jeitler\cmsAuthorMark{1}, A.~K\"{o}nig, N.~Krammer, I.~Kr\"{a}tschmer, D.~Liko, T.~Madlener, I.~Mikulec, E.~Pree, N.~Rad, H.~Rohringer, J.~Schieck\cmsAuthorMark{1}, R.~Sch\"{o}fbeck, M.~Spanring, D.~Spitzbart, W.~Waltenberger, J.~Wittmann, C.-E.~Wulz\cmsAuthorMark{1}, M.~Zarucki
% ...
\vskip\cmsinstskip
% ...
1:~Also at~Vienna~University~of~Technology,~Vienna,~Austria\\
% ...
\end{sloppypar}
\end{document}